\documentclass[8.5pt,twoside,twocolumn]{article}
\usepackage[super,sort&compress,comma]{natbib}
\usepackage{mhchem}
\usepackage{times,mathptmx}
\usepackage{sectsty}
\usepackage{balance}
\usepackage{amssymb}

\usepackage{graphicx} %eps figures can be used instead
\usepackage{lastpage}
\usepackage[format=plain,justification=raggedright,singlelinecheck=false,font=small,labelfont=bf,labelsep=space]{caption}
\usepackage{fancyhdr}
\usepackage[switch, modulo]{lineno}
\begin{document}
%\linenumbers

\thispagestyle{plain}

\twocolumn[
  \begin{@twocolumnfalse}
\noindent\LARGE{\textbf{A study of the influence of isotopic substitution on the melting point and temperature of maximum density of water by means of path integral simulations of rigid models.}}
\vspace{0.6cm}

\noindent\large{\textbf{Carl McBride,\textit{$^{a}$}, Juan L. Aragones\textit{$^{a}$},  Eva G. Noya,\textit{$^{b}$} and Carlos Vega\textit{$^{a\ddag}$}}}\vspace{0.5cm}

\noindent \textbf{\small{DOI:  10.1039/C2CP42393F}}
\vspace{0.6cm}

\noindent \normalsize{
The melting point of ice I$_{\mathrm{h}}$, as well as the temperature of maximum density (TMD) in the liquid phase,  
has been computed using the path integral Monte Carlo method. 
Two new models are introduced; TIP4PQ\_D2O and TIP4PQ\_T2O which are specifically designed to study D$_2$O and T$_2$O respectively. 
We have also used these models to study the ``competing quantum effects" proposal of Habershon, Markland and Manolopoulos;
the TIP4PQ/2005, TIP4PQ/2005 (D$_2$O) and  TIP4PQ/2005 (T$_2$O) models are able to study the isotopic substitution of hydrogen for deuterium or tritium whilst constraining the geometry, while the 
TIP4PQ\_D2O and TIP4PQ\_T2O models, where the O-H bond lengths are progressively shortened,  
permit the study of the influence of geometry (and thus dipole moment) on the isotopic effects.
For TIP4PQ\_D2O - TIP4PQ/2005 we found a melting point shift of 4.9 K (experimentally the value is 3.68K) and a TMD shift of 6K (experimentally 7.2K).
For TIP4PQ\_T2O - TIP4PQ/2005 we found a melting point shift of 5.2 K (experimentally the value is 4.49K) and a TMD shift of 7K (experimentally 9.4K).
}
\vspace{0.5cm}
 \end{@twocolumnfalse}
  ]
\footnotetext{\ddag~cvega@quim.ucm.es}
\footnotetext{\textit{$^{a}$~Departamento de Qu\'{\i}mica F\'{\i}sica, Facultad de Ciencias Qu\'{\i}micas, Universidad Complutense de Madrid, 28040 Madrid, Spain}}
\footnotetext{\textit{$^{b}$~Instituto de Qu\'{\i}mica F\'{\i}sica Rocasolano, Consejo Superior de Investigaciones Cient\'{\i}ficas, CSIC, Calle Serrano 119, 28006 Madrid, Spain}}
%%%%%%%%%%%%%%%%%%%%%%%%%%%%%%%%%%%%%%%%%%%%%%%%%%%%%%%%%%%%%%%%%%%%%
\section{Introduction}
%%%%%%%%%%%%%%%%%%%%%%%%%%%%%%%%%%%%%%%%%%%%%%%%%%%%%%%%%%%%%%%%%%%%%
Water molecules are composed of two hydrogen atoms and one oxygen atom (H$_2$O). 
One can substitute the hydrogen atom for isotopes of hydrogen: one can replace hydrogen for deuterium 
to form deuterium oxide ($^2$H$_2$O, or D$_2$O), popularly known as ``heavy water", or with tritium, forming tritium oxide ($^3$H$_2$O, or T$_2$O).
These isotopically substituted forms of water differ in their physical properties, such as heat capacities ($C_p$),
melting point ($T_m$), diffusion coefficients and the temperature of the maximum in density (TMD). 

Within the Born-Oppenheimer approximation the ``adiabatic surface", or potential energy surface (PES) is unaffected by 
such isotopic substitutions; any shift in  experimental
properties is due to a different probability  distribution of the different isotopes on the same PES,
something that classical statistical mechanics is unable to describe.
The quantum nature of the nuclei becomes increasingly  relevant when dealing with light atoms, 
in particular hydrogen. For this reason the incorporation of nuclear quantum effects in water 
is  germane.
A complete quantum mechanical description of water would require the solution of the electronic Schroedinger equation
for the electronic part of the Born-Oppenheimer approximation, in conjunction with, say, the path-integral 
formalism for the nuclear contribution  \cite{CPL_1985_117_0214,CPL_1984_103_0357,NATO_ASI_C_293_0155_photocopy}. Such a ``complete" solution is still in the far future \cite{chapter_Wong_2012}.
An intermediate approach is to use an empirical potential in place of the potential energy surface.  
A great many empirical potentials exist for water \cite{JML_2002_101_0219} (perhaps more than for any other molecule), having varying 
degrees of success \cite{PCCP_2011_13_19663}. One recently proposed classical model has been shown to 
be capable of reproducing a good number of the thermodynamic and transport properties of water, namely the TIP4P/2005 model \cite{JCP_2005_123_234505}.

Since in water nuclear quantum effects are significant \cite{JCP_2005_123_144506,JCP_2008_128_204107}, one must conclude that the 
parameters of these classical models implicitly include these quantum contributions.
A path integral simulation of a model such as TIP4P/2005 would be inappropriate
as it would lead to a ``double-counting" of the nuclear quantum effects, so in view of this 
a variant of this model was developed; namely the TIP4PQ/2005 model \cite{JCP_2009_131_024506}.
It was  found that  an increase of 0.02$e$ in the charge of the proton led to 
one of the most quantitative phase diagrams of water calculated to date \cite{PCCP_2012_14_10140}.

Recently a very interesting suggestion has been put forward by Habershon, Markland and Manolopoulos \cite{JCP_2009_131_024501},
that of ``competing quantum effects" in water;
they have proposed that zero point fluctuations lead to a longer O-H bond length, and thus a larger  dipole moment, making the 
water molecule ``less" quantum, whereas on the other hand inter-molecular quantum fluctuations serve to weaken the
hydrogen bonds, making the liquid as a whole more ``quantum".
An analogous process almost certainly takes place upon isotopic substitution;
the replacement of hydrogen with deuterium has two effects: on the one hand the hydrogen bond becomes stronger,
since D is less delocalised, i.e.  more classical, than H, 
whilst on the other hand replacing H with D reduces the intramolecular OH covalent bond length, which in turn 
decreases the dipole moment of the molecule, effectively reducing the strength of the hydrogen bond.

To study these competing quantum effects, and to
examine the influence of isotopic substitution on both the melting point of ice I$_{\mathrm{h}}$
and the location of the temperature of maximum density (TMD) in the liquid, we studied  the aforementioned TIP4PQ/2005 model along with two new models,
specifically designed
for simulations of D$_2$O and T$_2$O. Each of these models are both rigid and non-polarisable.
Water molecules are, beyond a doubt, flexible in nature.
Furthermore, it is  known that the isotopic substitutions noticeably affect the vibrational properties of water \cite{JPC_1996_100_01336} and also that there is a degree of
coupling between intermolecular and intramolecular modes.
Since intramolecular vibrations are generally high frequency oscillations a quantum rather
a classical description of these vibration would be desirable \cite{JCTC_2011_7_2903}.
That said however, the thermodynamic properties of the condensed phases of water are largely dominated by the intermolecular hydrogen bond rather than by the intramolecular vibrations. 
For this reason an analysis of 
how far one can go in the description of isotopic effects on the TMD and melting point of water using rigid models is still pertinent.
It will be shown that when the same model is used to study each of the  isotopes of water
the variation of the  properties, although qualitatively correct, are overestimated. 
However, it will be shown that by simply shortening the O-D and O-T bond length with respect to that of O-H,
then predictions of isotopic effects are in reasonable agreement with experimental results.
%%%%%%%%%%%%%%%%%%%%%%%%%%%%%%%%%%%%%%%%%%%%%%%%%%%%%%%%%%%%%%%%%%%%%
\section{Methodology and simulation details}
%%%%%%%%%%%%%%%%%%%%%%%%%%%%%%%%%%%%%%%%%%%%%%%%%%%%%%%%%%%%%%%%%%%%%
\begin{table*}[t]
\begin{center}
\caption{
Parameters for TIP4PQ/2005 and the new TIP4PQ\_D2O and TIP4PQ\_T2O models. The distance between the oxygen and hydrogen sites
is $d_{\mathrm{OH}}$. The angle, in degrees,  formed by hydrogen, oxygen, and the other hydrogen atom is
denoted by
H-O-H.  The Lennard-Jones site is located on the oxygen with parameters $\sigma$ and $\epsilon$.
The charge on the proton is $q_{\mathrm{H}}$. The negative charge is placed in a point M at a distance
$d_{\mathrm{OM}}$ from the oxygen along the H-O-H bisector.
}
\label{tip4p_2005_parameters}
\begin{tabular}{llllllll}
\hline
Model &$d_{\mathrm{OH}}$ (\AA) &\hspace{0.2cm} $\angle$H-O-H &\hspace{0.2cm} $\sigma$(\AA)&\hspace{0.2cm} $\epsilon/k_{B}$(K)& \hspace{0.2cm} $q_{\mathrm{H}}$(e)& \hspace{0.2cm}$d_{\mathrm{OM}}$(\AA) & $ p$ (Debye) \\
\hline
TIP4PQ/2005 & 0.9572 & \hspace{0.2cm}104.52 & \hspace{0.2cm}3.1589 & \hspace{0.2cm}93.2 &\hspace{0.2cm} 0.5764 &\hspace{0.2cm} 0.1546  & 2.388\\
TIP4PQ\_D2O & 0.9532 & \hspace{0.2cm}104.52 & \hspace{0.2cm}3.1589 & \hspace{0.2cm}93.2 &\hspace{0.2cm} 0.5764 &\hspace{0.2cm} 0.153954 &2.378\\
TIP4PQ\_T2O & 0.9512 & \hspace{0.2cm}104.52 & \hspace{0.2cm}3.1589 & \hspace{0.2cm}93.2 &\hspace{0.2cm} 0.5764 &\hspace{0.2cm} 0.1536   &2.373\\
\hline
\end{tabular}
\end{center}
\end{table*}
Recent experiments have indicated that the O-D bond length is shorter than the O-H bond length by $\approx 0.5\%$  \cite{PRL_2011_107_145501}.
In view of this we have taken the TIP4PQ/2005 model, and also shortened the O-D distance by a similar amount (by 0.004\AA  ~to be precise), resulting in the TIP4PQ\_D2O model.
The  location of the negative charge (situated on the massless site $M$) 
was also shifted so as to maintain the same ratio of  $d_{\mathrm{OM}}/d_{\mathrm{OH}}$ as in the original TIP4PQ/2005 model.
On doing this the relative distances between  the charges (responsible of the hydrogen bond strength)
and the Lennard-Jones site (which controls the short range repulsive forces) remains unchanged,  and provides a dipole to quadrupole ratio close to one, which has been shown to lead to a good phase diagram for TIP4P-like models \cite{PRL_2007_98_237801}.
We have also parameterised a model for  T$_2$O (TIP4PQ\_T2O) along the same lines. Given the paucity of experimental data for the O-T bond length in the liquid phase, we have taken the liberty of shortening the O-T bond length by
0.006\AA~ with respect to the H-O bond length, again maintaining the bond length ratio $d_{\mathrm{OM}}/d_{\mathrm{OH}}$. The resulting parameters are given in Table \ref{tip4p_2005_parameters}.
It is worth reiterating that all of these models are rigid and non-polarisable.
The path integral methodology for rigid rotors was employed and has been  discussed in detail elsewhere \cite{MP_2011_109_0149}, and we shall restrict ourselves to 
describing the most salient aspects of the work undertaken here.
The NVM propagator \cite{JCP_2011_134_054117}, exact for asymmetric tops, was used. The NVM propagator is based on the work of M{\"u}ser and  Berne for symmetric tops \cite{PRL_1996_77_002638}.
We used $P=7$ Trotter slices, or ``replicas",  for all simulations.
In Table \ref{table_ABC}  the rotational constants for the various models are presented.
Simulations of the liquid phase consisted of 360 molecules, and the ice  I$_{\mathrm {h}}$
phase consisted of 432 molecules. The proton disordered configuration of ice  I$_{\mathrm {h}}$, having both
zero dipole moment as well as  satisfying 
the Bernal-Fowler rules \cite{JCP_1933_01_00515}  was obtained by  
means of the algorithm of Buch  {\it et al.} \cite{JPCB_1998_102_08641}.
   The Lennard-Jones part of the potential was truncated at 8.5{\AA}
  and long range corrections were added. Coulombic interactions were treated using
  Ewald summation method.  
In the ice  I$_{\mathrm {h}}$ phase the $NpT$ ensemble was used, with anisotropic 
changes in the volume of the simulation box; each side being  able to fluctuate independently. 
All simulations were performed at a pressure of 1 bar.
  A Monte Carlo cycle consists of a trial move per particle (the number of particles is equal to $NP$ where
  $N$ is the number of water molecules) plus a trial volume change in the case of $NpT$ simulations.
\begin{table}[t]
\begin{tabular}{lrrr}
\hline
Model   & $A$ (cm$^{-1}$) & $B$ (cm$^{-1}$) & $C$ (cm$^{-1}$) \\
\hline
TIP4PQ/2005             &  27.432 & 14.595 & 9.526 \\
TIP4PQ/2005 (D$_2$O)    &  15.262 &  7.303 & 4.939 \\
TIP4PQ/2005 (T$_2$O)    &  11.211 &  4.877 & 3.398 \\
TIP4PQ\_D2O             &  15.390 &  7.365 & 4.981 \\
TIP4PQ\_T2O             &  11.353 &  4.939 & 3.441 \\
\hline
\end{tabular}
\caption{\label{table_ABC} Rotational constants for each of the models.}
\end{table}

%%%%%%%%%%%%%%%%%%%%%%%%%%%%%%%%%%%%%%%%%%%%%%%%%%%%%%%%%%%%%%%%%%%%%
\subsection{Calculation of the melting point}
%%%%%%%%%%%%%%%%%%%%%%%%%%%%%%%%%%%%%%%%%%%%%%%%%%%%%%%%%%%%%%%%%%%%%
Calculation of the melting point of ice  I$_{\mathrm {h}}$ consists of three 
steps:

Step 1: The first step is to calculate the classical melting point for the model of interest. This consists in
calculating the free energy of the solid phase via Einstein crystal calculations, followed by 
the addition of the residual entropy as calculated by Pauling \cite{JACS_1935_57_02680}. To obtain the free energy of the fluid phase
a thermodynamic path is constructed, making a 
connection to a Lennard-Jones reference fluid whose free energy is well known. Once the free energies of the
fluid and solid phases of the classical system are known for a reference state  thermodynamic
integration is performed to obtain 
the temperature for which both phases have the same chemical potential (at standard pressure),
i.e. the classical melting point. A thorough description of this procedure can be found in Ref. \cite{JPCM_2008_20_153101}.

Step 2:  At this classical melting point  one then calculates the difference in chemical potential
between ice  I$_{\mathrm {h}}$ and water for the quantum system ($\Delta \mu$) via:
\begin{eqnarray}
\frac{\Delta \mu} {k_BT} =   
\int_{0}^{1}\frac{1}{\lambda'}
\left[ \left\langle \frac{K_{\mathrm{Ih}}}{Nk_BT} \right\rangle -  \left\langle \frac{K_{\mathrm{liquid}}}{Nk_BT} \right\rangle
 \right] \mathrm{d}\lambda'
\label{eq_guay}
\end{eqnarray}
where $K$ represents the total kinetic energy, given by:
\begin{equation}
K = K_{tra} + K_{rot},
\end{equation}
where
\begin{equation}
\label{Ktra_eq}
K_{tra} = \frac{3NP}{2\beta} - \left\langle  \frac{MP}{2\beta^2\hbar^2} \sum_{i=1}^{N} \sum_{t=1}^{P} ({\bf r}_{i}^{t}-{\bf r}_{i}^{t+1})^{2} \right\rangle_{NpT},
\end{equation}
where $P$ is the number of `beads" and $M$ is the molecular mass, and
\begin{equation}
K_{rot}=\left\langle \frac{1}{P}  \sum_{t=1}^P  \sum_{i=1}^N e_{rot,i}^{t,t+1}
\right\rangle_{NpT}
\end{equation}
where $e_{rot}^{t,t+1}$ is the rotational energy term of the NVM propagator
(for details see Ref.  \cite{JCP_2011_134_054117}).
The parameter $\lambda'$ is defined so that the mass of each atom $i$ of the system 
is scaled as $m_i= m_{i,0}/\lambda'$ where $m_{i,0}$ is the mass of atom $i$ in the original 
system. The values of  $m_{i,0}$ for H, D, T and O were taken from Ref. \cite{RMP_2008_80_00633}.  
On increasing the atomic masses by the factor $1/\lambda'$ the geometry and centre of mass of the molecule remains unchanged.
Similarly the eigenvalues of the inertia tensor,  and thus the energies of the asymmetric top
appearing in the rotational propagator, are also scaled by this factor. 
Such a linear scaling particularly convenient for practical purposes.
However, the same is not true for a transformation from, say, the TIP4PQ/2005 to TIP4PQ\_D2O models, 
since the geometry and mass distribution  varies  between the models.
Similarly there is no  simple scaling for the values of the rotational constants $A$, $B$ and $C$ used in the calculation of the propagator.
For this reason we perform the integration to infinite mass for each of the models, rather than perform a transformation between models.
Eq. \ref{eq_guay} embodies the idea that 
the phase that has the  higher kinetic energy will also have the higher 
chemical potential and as a result will become less stable in the quantum system.
Note  that for the TIP4PQ/2005 model the melting point of the classical system is the same for H$_2$O, D$_2$O and T$_2$O
since the melting point of a classical system is independent of the molecular mass.
The integrand of Eq. \ref{eq_guay} represents the transformation from H$_2$O (or D$_2$O or T$_2$O) to an infinitely massive 
molecule of water.
This integral was evaluated using seven values of $\lambda'$
between 1/7 and 1 (i.e. $\lambda'=1,6/7,5/7,4/7,3/7,2/7$, and $1/7$) 
by performing runs of about one million cycles at each value of $\lambda'$.
We do not go beyond $\lambda'=1/7$  due to the increased expense in the evaluation of the propagator, 
however, in Ref. \cite{PCCP_2012_14_10140} it is shown that the integral is well behaved down to $\lambda=0$ for the case of the harmonic oscillator.
Furthermore our direct coexistence
simulations \cite{JCP_2006_124_144506,PCCP_2012_14_10140} corroborate our melting point calculations, which take to be an indication that there is no ``anomalous" behaviour in the region between $\lambda'=1/7$ and $\lambda'=0$.

Step 3: Again using thermodynamic integration \cite{JPCM_2008_20_153101},
the free energy of each phase of the quantum system is determined as a function 
of $T$: 
\begin{equation}
\frac{G(T_2,p)}{Nk_BT_2} = \frac{G(T_1,p)}{Nk_BT_1} - \int_{T_1}^{T_2} \frac{H(T)}{Nk_BT^2} ~\mathrm{d}T 
\label{eq_inegrate_pressure}
\end{equation}
where $G$ is the Gibbs energy function and $H$ is the enthalpy.
This provides the location of the melting point of the quantum system as the temperature at which 
the chemical potential of ice  I$_{\mathrm {h}}$ and water become identical. 
%%%%%%%%%%%%%%%%%%%%%%%%%%%%%%%%%%%%%%%%%%%%%%%%%%%%%%%%%%%%%%%%%%%%%
\subsection{Calculation of the TMD}
%%%%%%%%%%%%%%%%%%%%%%%%%%%%%%%%%%%%%%%%%%%%%%%%%%%%%%%%%%%%%%%%%%%%%
The determination of the location of the TMD consisted in particularly long simulation runs
(up to 9 million Monte Carlo cycles per temperature) for a range of temperatures that bracket the location of the TMD. Once the densities as a function of temperature were obtained, they were fitted to  a quadratic polynomial,
whose maxima was taken to be the location of the TMD.

%%%%%%%%%%%%%%%%%%%%%%%%%%%%%%%%%%%%%%%%%%%%%%%%%%%%%%%%%%%%%%%%%%%%%
\section{Results}
%%%%%%%%%%%%%%%%%%%%%%%%%%%%%%%%%%%%%%%%%%%%%%%%%%%%%%%%%%%%%%%%%%%%%
\subsection{Melting point of the TIP4PQ/20005 model}
\begin{figure}[t]
\begin{center}
\includegraphics[angle=270,width=87mm,clip]{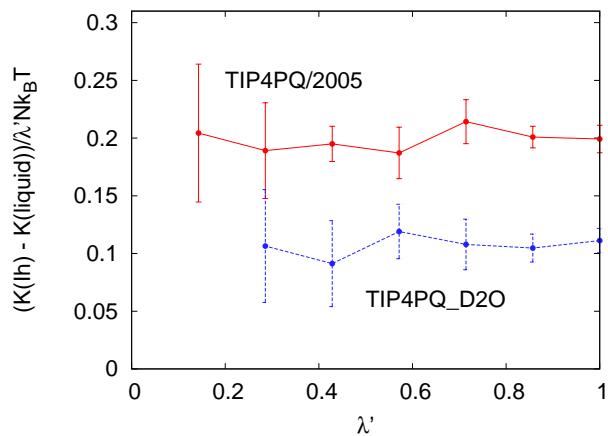}
\caption{\label{fig_integrand}Integrand of Eq. \ref{eq_guay}
 (i.e. $(K_{\mathrm{Ih}}- K_{\mathrm{liquid}})/(\lambda'Nk_BT)$) as a function of $\lambda'$.
The integral of the curves (from 0 to 1) yields  $\Delta \mu$(I$_{\mathrm {h}}$ - liquid)/  (k$_B$T).
Key: TIP4PQ/2005 red line, TIP4PQ\_D2O blue dashed line. 
%TIP4PQ/2005 (D$_2$O) is (xxx), TIP4PQ/2005 (T$_2$O) is (xxx), TIP4PQ\_D2O is (xxx) and TIP4PQ\_T2O is (xxx).
}
\end{center}
\end{figure}

%%%%%%%%%%%%%%%%%%%%%%%%%%%%%%%%%%%%%%%%%%%%%%%%%%%%%%%%%%%%%%%%%%%%%
The classical value of the melting point for the TIP4PQ/20005 was calculated to be 282K \cite{PCCP_2012_14_10140}.
As per Step 2 of the methodology outlined previously, the integrand of equation 
\ref{eq_guay} was calculated  and the results are presented in Figure \ref{fig_integrand}.

One can see that this integrand is positive, indicating that the molecules have 
more kinetic energy in the ice phase than they do in the liquid phase, thus ice  I$_{\mathrm {h}}$ is less stable in the quantum system, which in turn implies 
that the melting point will move to lower temperatures in the quantum system. 
Since the hydrogen bonds are stronger in the ice phase quantum effects are more influential in the ice phase. 
The integrand is fairly smooth, and so can be fitted to a straight line
down to very small values of  $\lambda'$, and it is from this fit that we obtain the value of the integral.
The values for these integrals are presented in Table \ref{table_integrals}.
\begin{table*}[t]
\begin{tabular}{lcc}
\hline
Model   & T (K) & $\Delta \mu$(I$_{\mathrm {h}}$ - liquid) /  (k$_B$T)  \\% &   integral $(H/Nk_BT^2)$ liquid & ice\\
\hline
TIP4PQ/2005            & 282 &  0.198  \\%   &  -2.3158   &  -2.5824 \\
TIP4PQ/2005 (D$_2$O)   & 282 &  0.120  \\%   &  -2.4704   &  -2.7360  \\
TIP4PQ/2005 (T$_2$O)   & 282 &  0.092  \\%   &  -2.5248   &  -2.8096  \\
TIP4PQ\_D2O            & 276  &  0.108  \\%   &  -2.0098   &  -2.2282  \\
TIP4PQ\_T2O            & 273  &  0.084  \\%   &  -1.8203   &  -2.0208  \\
\hline
\end{tabular}
\caption{\label{table_integrals} The difference in the chemical potential between the ice and liquid phases in the
quantum system, evaluated at the $T_m$ of the classical system at a pressure of $p=1$ bar. (Error $\pm$ 0.01).}
\end{table*}
Having these integrals we proceeded to Step 3, i.e. the thermodynamic integration given in Eq.  \ref{eq_inegrate_pressure}.
To do this path integral simulations were performed for both ice  I$_{\mathrm {h}}$  and water at various temperatures along the $p=1$ bar isobar.
The melting point of the quantum system  is the temperature at which the chemical potential  of both  I$_{\mathrm {h}}$  and water are the same.
The resulting melting points are given in Table \ref{table_Tm}.
\begin{table*}[t]
\begin{tabular}{lrcrc}
\hline
Model   & $T_m$   & $ T_m - T_m^{H_2O}$ &      $T_{ {\mathrm{TMD}} }$    &  $T_{TMD} - T_{TMD}^{H_2O}$  \\
\hline
H$_2$O (experiment)          &  273.15   &   0   &    277.13    &  0 \\
D$_2$O (experiment)         &  276.83   &  3.68    &    284.34  &  7.2 \\
T$_2$O (experiment)         &  277.64   &  4.49    &     286.55 &  9.4 \\
\hline
  & $T_m$   & $ T_m - T_m^{TIP4PQ/2005}$ &      $T_{ {\mathrm{TMD}} }$    &  $T_{TMD} - T_{TMD}^{TIP4PQ/2005}$  \\
\hline
TIP4PQ/2005             &  258.3               &   0   &   284  &  0 \\
TIP4PQ/2005 (D$_2$O)    &  267.7 &  9.4  &    295          & 11 \\
TIP4PQ/2005 (T$_2$O)    &  271.8 & 13.5  &            300  & 16 \\
\hline
TIP4PQ/2005             &  258.3               &   0   &   284  &  0 \\
TIP4PQ\_D2O             &  263.2 &  4.9  &   290 & 6 \\
TIP4PQ\_T2O             &  263.5 & 5.2 &   291   & 7 \\
\hline
\end{tabular}
\caption{\label{table_Tm} Melting points and temperatures of maximum density of the models (all temperatures are in Kelvin).}
\end{table*}

The melting point of the TIP4PQ/2005 model is 258K, which is approximately 15K below the experimental value.
TIP4PQ/2005 is not alone in underestimating the melting point; the flexible q-TIP4P/F model \cite{JCP_2009_131_024501}, also designed for 
use in path integral simulations, 
has a similar melting point (251 K \cite{PCCP_2011_13_19714}).
This is probably due to the fact that both the q-TIP4P/F and TIP4PQ/2005 models are derived from the classical TIP4P/2005 model
which has $T_m=252$ K.
The TTM2.1-F and TTM3-F models \cite{JCP_2008_128_074506}, 
which are both flexible and polarisable and were obtained from fits
to high level {\it ab initio} calculations, 
have somewhat lower melting points; 228 K \cite{JPCC_2008_112_00324}  and 225 K \cite{JPCL_2010_001_02316}  respectively,
while the q-SPC/Fw model \cite{JCP_2006_125_184507} has a $T_m$ of 195 K \cite{JCP_2009_131_024501}.  
Conversely, density functional theory predictions for the melting point tend to significantly overestimate the experimental value;
two common functionals, PBE0 and BLY3P, \cite{JCP_2009_130_221102} have a melting point of  $T_m=415$ K.
 
From Table \ref{table_deltaH} one can  see that the TIP4PQ/2005 models  underestimate the melting enthalpy (1.099 kcal/mol, whereas the 
experimental value is 1.436 kcal/mol). The enthalpy of melting was obtained from $NpT$ simulations of both the solid phase and the liquid phase, for each model, both at the melting point, then simply taking the
enthalpy difference at this temperature.
Both of these results, the melting point and the melting enthalpy, were also underestimated 
in classical simulations of the classical model TIP4P/2005 (251 K and 1.15 kcal/mol respectively). From this we can deduce that 
the inclusion of nuclear quantum effects has relatively little influence over these properties, and that any discrepancy with experiment
is due to the approximate description of the PES implied in the empirical TIP4P/2005 and TIP4PQ/2005 models.
% As to the performance of other models the melting enthalpy of TTM2.1-F was reported to be 1.00 kcal/mol\cite{JPCC_2008_112_00324}.

%%%%%%%%%%%%%%%%%%%%%%%%%%%%%%%%%%%%%%%%%%%%%%%%%%%%%%%%%%%%%%%%%%%%%
\subsection{Isotope effects on the melting point }
%%%%%%%%%%%%%%%%%%%%%%%%%%%%%%%%%%%%%%%%%%%%%%%%%%%%%%%%%%%%%%%%%%%%%
In classical simulations the melting point is independent of the molecular mass, thus the 
melting points of the  TIP4PQ/2005 (D$_2$O) and  TIP4PQ/2005 (T$_2$O) models is the same as that of the 
TIP4PQ/2005 model, namely 282K. As per Step 2 of the 
methodology outlined previously the  integrand of equation
\ref{eq_guay} was calculated (see  Figure \ref{fig_integrand}) and the integral of  Eq.(1) evaluated
(see Table  \ref{table_integrals}).
Thermodynamic integration was then undertaken leading to the melting points of the quantum system;  
268K for TIP4PQ/2005 (D$_2$O) and 272K for TIP4PQ/2005 (T$_2$O). This increase in the melting point qualitatively mirrors experimental results,
however, the magnitude of the shift is over estimated (see column 3 of Table 4).

\begin{table*}[t]
\begin{tabular}{lrrr}
\hline
Model   &   $\Delta H (T_m)$ (kcal/mol)  & $\rho_{ {\mathrm {Ih}} }$ ($T_m$)  & $\rho_{ {\mathrm {liquid}} }$ ($T_m$)\\
\hline
H$_2$O (experiment)       & 1.436 & 0.917   &   0.999\\  % (-42846.71 + 47428.48) /4184 (kcal/mol) a 260K
D$_2$O (experiment)       & 1.509 & 1.018   &   1.105\\  % (-42846.71 + 47428.48) /4184 (kcal/mol) a 260K
T$_2$O (experiment)       & ---  & ---   &   ---\\  % (-42846.71 + 47428.48) /4184 (kcal/mol) a 260K
TIP4PQ/2005              & 1.099 & 0.919   &  0.988 \\  % (-42869.3417 +47468.0202)/4184
TIP4PQ/2005 (D$_2$O)    & 1.189  & 1.028   &  1.103 \\ % (-44751.35 + 49755.53 ) / 4184   a 270 K  
TIP4PQ/2005 (T$_2$O)    & 1.285  & 1.134   &  1.212\\ % (-45659.57 + 51248.38) / 4184
TIP4PQ\_D2O             & 1.133  & 1.024 &   1.091\\
% (-10.6702227400403) + ((22.0270086 / 1.21360952253739) / 41 840) = -10.6697889 
%(-11.8624825645542) + ((22.0270086 / 1.12778568637326) / 41 840) = -11.8620158
% (-10.6697889) + 11.8620158 = 1.1922269
TIP4PQ\_T2O             & 1.192  & 1.128   & 1.214  \\
\hline
\end{tabular}
\caption{\label{table_deltaH} The change in enthalpy at the melting points of the models along with densities in units of g/cm$^{3}$ (experimental values from IAPWS-95/NIST Standard Reference Data).}
\end{table*}
The same procedure was applied to the TIP4PQ\_D2O and TIP4PQ\_T2O models, which have classical melting points of 276 K and 273 K respectively.
We find a difference of 4.9 K between TIP4PQ\_D2O and TIP4PQ/2005, and 5.2 K between TIP4PQ\_T2O and TIP4PQ/2005,
which is in much better agreement with the experimental value of the shift. 
Similar values for the melting point differences were found for the q-TIP4P/F model; 6.5K between D$_2$O and H$_2$O, and 8.2 for T$_2$O with respect to H$_2$O \cite{JCP_2010_133_144511}.
For D$_2$O the melting enthalpy is found to be from experiments about 0.07 kcal/mol higher than that of water, 
whereas TIP4PQ/2005 predicts an increase of 0.09 kcal/mol and TIP4PQ\_D2O predicts an increase of about 0.04 kcal/mol.

%%%%%%%%%%%%%%%%%%%%%%%%%%%%%%%%%%%%%%%%%%%%%%%%%%%%%%%%%%%%%%%%%%%%%
\subsection{Isotope effects on the temperature of maximum density (TMD) }
\begin{figure}[t]
\begin{center}
\includegraphics[angle=0,width=87mm,clip]{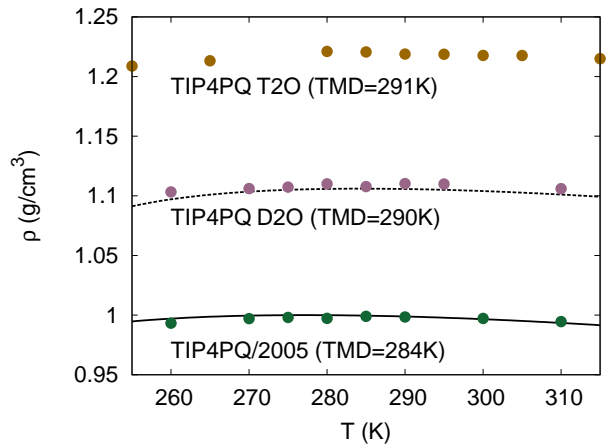}
\caption{\label{fig_TMD} Plot of the isobars ($p=1$ bar) used to calculate the TMD (points), along with fits to experimental results \cite{AE_1966_20_168} for H$_2$O and D$_2$O (lines) (the authors were unable to locate experimental data for T$_2$O).}
\label{fig_tmd}
\end{center}
\end{figure}

%%%%%%%%%%%%%%%%%%%%%%%%%%%%%%%%%%%%%%%%%%%%%%%%%%%%%%%%%%%%%%%%%%%%%
Experimentally deuteration of water shifts the TMD by 7.2 K \cite{JPCRD_1982_11_0001}, and tritiation by 9.4 K \cite{book_FFranks_Water_Matrix_Life}.
Our previous results \cite{JCP_2009_131_124518} indicate that deuteration and tritiation of the TIP4PQ/2005 model
tended to overestimate this shift. Our new models now slightly underestimate this shift (see Table \ref{table_Tm} and figure \ref{fig_tmd}).
In view of the fact that the error  bar for the TMD is fairly large ($\pm 2$K), these results are reasonable.  
Especially when one bears in mind that an isotopic shift in the TMD is not always present in a number of recent models \cite{JCP_2010_133_144511,JCP_2007_127_074506}, as is the case of the 
q-TIP4P/F \cite{JCP_2009_131_024501} and the TTM2.1-F models.

%%%%%%%%%%%%%%%%%%%%%%%%%%%%%%%%%%%%%%%%%%%%%%%%%%%%%%%%%%%%%%%%%%%%%
\subsection{The temperature difference between the melting point and the TMD}
%%%%%%%%%%%%%%%%%%%%%%%%%%%%%%%%%%%%%%%%%%%%%%%%%%%%%%%%%%%%%%%%%%%%%
 Of particular interest is the difference between the TMD and $T_m$. Experimentally this difference is 3.98 K for H$_2$O, 7.5 K for D$_2$O, and 8.9 K
for T$_2$O. From our simulations we obtain 25.7 K for TIP4PQ/2005, 26.8 K for TIP4PQ\_D2O and 27.5 K for TIP4PQ\_T2O.
As can be seen all the models presented in this work are unable to describe the difference between the temperature
of the TMD and the melting point temperature. Thus the inclusion of nuclear quantum effects does not solve the disagreement 
with experiment indicating that the origin of this failure is the approximate character of the PES. The same is true for the q-TIP4P/F  which predicts a difference between 
the TMD  and the melting point of 26 K (the model has the TMD at 277 K and the melting point at 251 K). For the  TTM2.1-F models \cite{JPC_A_2006_110_004100} the difference between the 
TMD and the melting point is even higher (45 K) (the melting point is located at 228 K \cite{JPCC_2008_112_00324} and the 
TMD at 273 K \cite{JCP_2008_128_074506}) . One can conclude that no model designed for path integral simulations thus far is able to reproduce
the difference between the TMD and the melting point found experimentally.

%%%%%%%%%%%%%%%%%%%%%%%%%%%%%%%%%%%%%%%%%%%%%%%%%%%%%%%%%%%%%%%%%%%%%
\subsection{Isotope effects on molar volumes} 
%%%%%%%%%%%%%%%%%%%%%%%%%%%%%%%%%%%%%%%%%%%%%%%%%%%%%%%%%%%%%%%%%%%%%
Bridgman described that the ``{\it molar volume of D$_2$O is always greater than that of H$_2$O at the same pressure and temperature}" \cite{JCP_1935_03_00597}.
Experimentally for ice  I$_{\mathrm {h}}$ this difference was seen to be of the order of 
0.2\% at 220K \cite{ACSB_1994_50_644,ACSB_2012_68_091}.
From our simulations of ice I$_{\mathrm {h}}$  we obtained 32.410  \AA$^3$/molecule for TIP4PQ/2005  at 220K, and  32.303  \AA$^3$/molecule for TIP4PQ\_D2O, also  at 220K,
which is similar to experimental results of 32.367 and 32.429 for H$_2$O and D$_2$O respectively \cite{PRL_2012_108_193003}, also at 220K.
From this one can see that the models used here 
are unable to capture this (albeit subtle) effect.
However, recent {\it ab initio} density functional theory calculations have been able to reproduce this effect  \cite{PRL_2012_108_193003},
although there is a $\approx 4\%$ error in the densities themselves.
It would be interesting to see  whether  {\it ab initio} density functional theory calculations are also capable of reproducing
the isotopic shifts found in the $T_m$ and the TMD.

%%%%%%%%%%%%%%%%%%%%%%%%%%%%%%%%%%%%%%%%%%%%%%%%%%%%%%%%%%%%%%%%%%%%%
\subsection{Isotope effects on the structure}
%%%%%%%%%%%%%%%%%%%%%%%%%%%%%%%%%%%%%%%%%%%%%%%%%%%%%%%%%%%%%%%%%%%%%
\begin{figure}[t]
\begin{center}
\includegraphics[angle=0,width=87mm,clip]{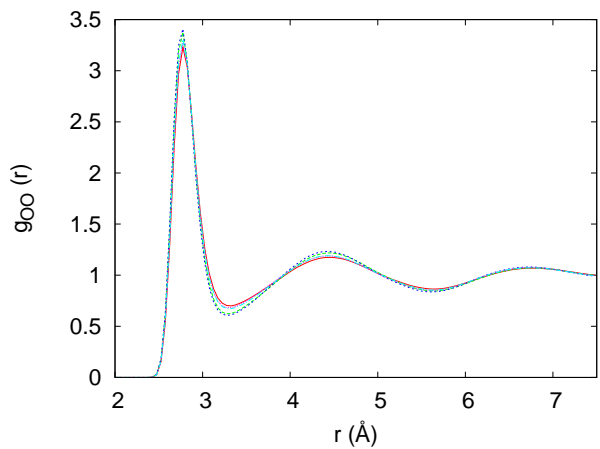}
\caption{\label{fig_rdf_OO} Plot of the O-O radial distribution function  for water at 290 K and $p=1$ bar. Key: TIP4PQ/2005  red line,  TIP4PQ/2005 (D$_2$O) dashed green line, TIP4PQ/2005 (T$_2$O) dashed blue line, TIP4PQ\_D2O dotted pink line and TIP4PQ\_T2O dot-dashed cyan line.}
\end{center}
\end{figure}
\begin{figure}[t]
\begin{center}
\includegraphics[angle=0,width=87mm,clip]{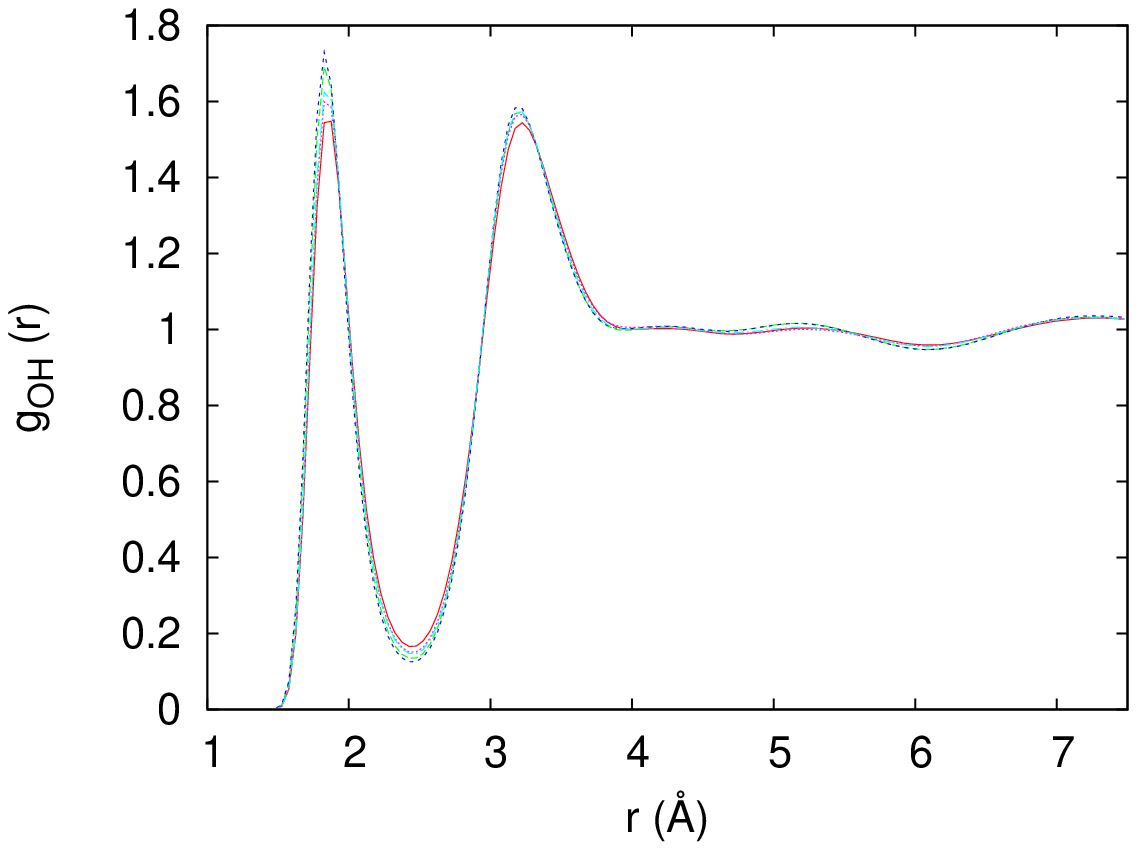}
\caption{\label{fig_rdf_OH} Plot of the O-H radial distribution function  for water at 290 K and $p=1$ bar. The same key as in Fig. \ref{fig_rdf_OO}.}
\end{center}
\end{figure}
\begin{figure}[t]
\begin{center}
\includegraphics[angle=0,width=87mm,clip]{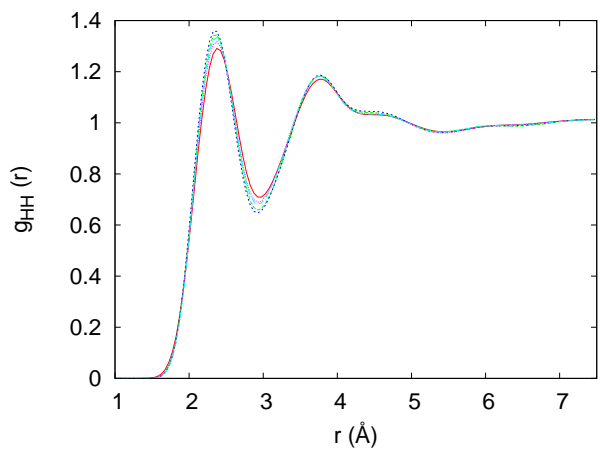}
\caption{\label{fig_rdf_HH} Plot of the H-H radial distribution function  for water at 290 K and $p=1$ bar. The same key as in Fig. \ref{fig_rdf_OO}. }
\end{center}
\end{figure}
In Figures \ref{fig_rdf_OO}-\ref{fig_rdf_HH} we provide the atom-atom distribution functions for O-O, H-H and O-H for each of the models studied.
From these plots, whose salient features are compiled in Table \ref{table_OOrdf}, one can observe that 
the structure of the new models, TIP4PQ\_D2O and TIP4PQ\_T2O is very similar to that of TIP4PQ/2005, more so than 
that of TIP4PQ/2005 (D$_2$O) and TIP4PQ/2005 (T$_2$O). This implies that the structure of the fluid phase of isotopically 
substituted water is almost indistinguishable from that of H$_2$O, an assumption that experimentalists often make,
and one that seems to be justified by our simulation results.
\begin{table*}
\begin{center}
\caption{\label{table_OOrdf}O-O, O-H and H-H radial distribution function of water for the various models at 290 K and $p=1$ bar.}
\begin{tabular}{l|ll|ll}
\hline
Model     &  \multicolumn{2}{l}{peak 1} &  \multicolumn{2}{l}{peak 2}       \\  \cline{2-5}
          &  $r$ (\AA)  &  height          &  $r$ (\AA)  &  height                                      \\
\hline
TIP4PQ/2005        &            2.775   & 3.23 &    4.475   & 1.17   \\
TIP4PQ/2005 (D$_2$O)       &    2.775   & 3.38 &    4.425   & 1.22   \\
TIP4PQ/2005 (T$_2$O)       &    2.775   & 3.41 &    4.425   & 1.23   \\
TIP4PQ\_D2O        &            2.775   & 3.27 &    4.425   & 1.19   \\
TIP4PQ\_T2O        &            2.775   & 3.30 &    4.425   & 1.19   \\
\hline
TIP4PQ/2005        &            1.875   & 1.55 &    3.225   & 1.54   \\
TIP4PQ/2005 (D$_2$O)       &    1.825   & 1.69 &    3.225   & 1.57   \\
TIP4PQ/2005 (T$_2$O)       &    1.825   & 1.73 &    3.175   & 1.58   \\
TIP4PQ\_D2O        &            1.825   & 1.60 &    3.225   & 1.57   \\
TIP4PQ\_T2O        &            1.825   & 1.62 &    3.225   & 1.57   \\
\hline
TIP4PQ/2005        &            2.375   & 1.29 &    3.775   & 1.17   \\
TIP4PQ/2005 (D$_2$O)       &    2.375   & 1.34 &    3.775   & 1.18   \\
TIP4PQ/2005 (T$_2$O)       &    2.375   & 1.36 &    3.775   & 1.19   \\
TIP4PQ\_D2O        &            2.375   & 1.32 &    3.775   & 1.18   \\
TIP4PQ\_T2O        &            2.375   & 1.33 &    3.775   & 1.18   \\
\hline
\end{tabular}
\end{center}
\end{table*}
%%%%%%%%%%%%%%%%%%%%%%%%%%%%%%%%%%%%%%%%%%%%%%%%%%%%%%%%%%%%%%%%%%%%%
\subsection{Isotope effects on the diffusion coefficient}
%%%%%%%%%%%%%%%%%%%%%%%%%%%%%%%%%%%%%%%%%%%%%%%%%%%%%%%%%%%%%%%%%%%%%
In the work of Habershon et al. \cite{JCP_2009_131_024501} a H$_2$O/D$_2$O 
diffusion coefficient ratio of 1.28 was found for the  q-TIP4P/F model at 298K (experimentally it is 1.30). 
Although it is not possible to directly compute the diffusion coefficient from Monte Carlo runs, a rough estimate can be obtained
by calculating the mean square displacement of the molecules after a fixed number of Monte Carlo cycles. 
In our simulations the ratio of the mean square displacement (after 200,000 MC cycles) 
between TIP4PQ/2005 and TIP4PQ/2005 (D$_2$O), was 1.33.
When we compare TIP4PQ/2005 with TIP4PQ\_D2O we obtain a ratio of 1.17, indicating that the new model decreases the
differences between D$_2$O and H$_2$O, 
in line with the results for the radial distribution function.
%%%%%%%%%%%%%%%%%%%%%%%%%%%%%%%%%%%%%%%%%%%%%%%%%%%%%%%%%%%%%%%%%%%%%
\subsection{Heat capacity $C_p$}
%%%%%%%%%%%%%%%%%%%%%%%%%%%%%%%%%%%%%%%%%%%%%%%%%%%%%%%%%%%%%%%%%%%%%
The isobaric heat capacity was obtained from a differential of the enthalpy with respect to temperature.
At 280K  the values we obtained were  (in cal/mol/K)
TIP4PQ/2005 17.4, TIP4PQ\_D2O 18.7 and for TIP4PQ\_T2O 19.5 
%and TIP4PQ/2005 (T$_2$O) is XXX. 
These results 
compare favourably with the experimental values;  H$_2$O is 17.9 \cite{JPC_1982_86_00998}  and    D$_2$O is 20.3 \cite{TA_1996_286_051} .
%%%%%%%%%%%%%%%%%%%%%%%%%%%%%%%%%%%%%%%%%%%%%%%%%%%%%%%%%%%%%%%%%%%%%
\section{Conclusions}
%%%%%%%%%%%%%%%%%%%%%%%%%%%%%%%%%%%%%%%%%%%%%%%%%%%%%%%%%%%%%%%%%%%%%
We have seen the ``competing quantum effects" interpretation of Habershon et al. in action;
the TIP4PQ/2005, TIP4PQ/2005 (D$_2$O) and TIP4PQ/2005 (T$_2$O) models all have the same geometry and charge distribution, 
thus they all have the same dipole moment. When one examines the radial distribution functions
one can see that the TIP4PQ/2005 (D$_2$O) and TIP4PQ/2005 (T$_2$O) models have stronger 
features than the TIP4PQ\_D2O and TIP4PQ\_T2O models, whose dipole moments are smaller.
The new models presented here for D$_2$O and T$_2$O were designed by shortening the O-H bond length in line with the values 
presented by Zeidler et al.  \cite{PRL_2011_107_145501}.
It is worth noting that a bond length reduction by as much as 4\%, as suggested by Soper and Benmore \cite{PRL_2008_101_065502},
would probably  have led to a significant error in our evaluation the isotopic influence on the melting point.
We have seen that the models studied  in this work underestimate the melting point by $\approx 7\%$
and the melting enthalpy by  $\approx 30\%$.
This is almost certainly due to 
the approximate nature of the empirical potentials used here, failing to reproduce the experimental PES.
The situation is more favourable when one considers isotopic shifts. In general we have seen that 
the models qualitatively reproduce the experimental trends. The models proposed in this work for D$_2$O and T$_2$O predict
an increase both in the TMD and in the melting point which are more realistic than those predicted by isotopic substitution in the TIP4PQ/2005 models. 

This work was funded by grants FIS2010-16159 and FIS2010-15502 of the  Direcci\'on General de Investigaci\'on
and  S2009/ESP-1691-QF-UCM (MODELICO) of the Comunidad Aut\'onoma de Madrid. One of the authors (J. L. A.)
would like to thank the MEC for a FPI pre-doctoral studentship.
The authors would like to thank Prof. Jose Luis Abascal for many helpful discussions.

\footnotesize{
%\bibliography{./bibliography} %your .bib file
\bibliographystyle{rsc} %the RSC's .bst file

\providecommand*{\mcitethebibliography}{\thebibliography}
\csname @ifundefined\endcsname{endmcitethebibliography}
{\let\endmcitethebibliography\endthebibliography}{}

}

\end{document}